\documentclass[preprint,aps,prd,showpacs,nofootinbib]{revtex4}
\usepackage{mathrsfs}
\usepackage{amsmath}
\usepackage{graphicx}
\usepackage{subfigure}

\newcommand{\bea}{\begin{eqnarray}}
\newcommand{\eea}{\end{eqnarray}}
\newcommand{\beq}{\begin{equation}}
\newcommand{\eeq}{\end{equation}}

\def\/{\over}

\begin{document}

\title{Spontaneous excitation of a static multilevel atom coupled with electromagnetic vacuum fluctuations in
Schwarzschild spacetime}
\author{Wenting Zhou$^{1}$ and Hongwei Yu$^{1,2,}$\footnote{Corresponding author}
}
\affiliation{$^1$  Institute of Physics and Key Laboratory of Low
Dimensional Quantum Structures and Quantum
Control of Ministry of Education,
Hunan Normal University, Changsha, Hunan 410081, China \\
$^2$ Center for Nonlinear Science and Department of Physics, Ningbo
University, Ningbo, Zhejiang 315211, China
}

\begin{abstract}
We study the spontaneous excitation of a radially polarized static
multilevel atom outside a spherically symmetric black hole in
multi-polar interaction with quantum electromagnetic fluctuations in
the Boulware, Unruh and Hartle-Hawking vacuum states. We find that
spontaneous excitation does not occur in the Boulware vacuum,  and,
in contrast to the scalar field case, spontaneous emission rate is
not well-behaved at the event horizon as result of the blow-up of
the proper acceleration of the static atom. However, spontaneous
excitation can take place both in the Unruh and the Hartle-Hawking
vacua as if there were thermal radiation from the black hole.
Distinctive features in contrast to the scalar field case are the
existence of a term proportional to the proper acceleration squared
in the rate of change of the mean atomic energy in  the Unruh and
the Hartle-Hawking vacuums and the structural similarity in the
spontaneous excitation rate between the static atoms outside a black
hole and uniformly accelerated ones in a flat space with a
reflecting boundary, which is particularly dramatic at the event
horizon where a complete equivalence exists.

\end{abstract}
 \maketitle

\baselineskip=16pt

\section{Introduction}

It is well-known that spontaneous emission (and excitation), as one
of the most important features of atoms, can be attributed to
vacuum fluctuations \cite{Welton48,GRF83}, or radiation reaction
\cite{Ackerhalt}, or a combination of them \cite{P.W.Milonni}. The
ambiguity in theoretical interpretation, which roots in the freedom
in the choice of ordering of commuting operators of the atom and
field in a Heisenberg picture approach to the problem, was resolved
by Dalibard, Dupont- Roc and Cohen-Tannoudji
(DDC)~\cite{DDC82,DDC84} who proposed a formalism that demands a
symmetric operator ordering so that the contributions of vacuum
fluctuations and radiation reaction can be distinctively separated.
 The DDC proposal successfully resolves the problem of
the stability of an inertial atom in its ground state in vacuum as a
result of the delicate balance between the contributions of vacuum
fluctuations and radiation reaction to the rate of change of the
mean atomic energy. Recent investigations using the DDC formalism on
the excitation of uniformly accelerated two-level atoms in
interaction with fluctuating quantized massless scalar fields in
vacuum in a flat spacetime with~\cite{YuLu05} and without
boundaries~\cite{Audretsch94} and that of static atoms outside a
Schwarzschild black hole~\cite{ZhouYu07} show that the delicate
balance no longer exists in the cases under consideration, thus
making the transition of the atom from ground state to excited
states possible, i.e., excitation of atoms spontaneously occurs. The
spontaneous excitation of uniformly accelerated atoms in the flat
spacetime can be regarded as providing a physically appealing
interpretation of the Unruh effect~\cite{Unruh76}, since the
spontaneous excitation of accelerated atoms gives a physically
transparent illustration for why an accelerated detector clicks,
while the spontaneous excitation of the static atoms outside a
Schwarzschild black hole can be considered as providing another
approach to the derivation of the Hawking radiation  and it shows
pleasing consistence of two different physical phenomena, the
Hawking radiation and the spontaneous excitation of atoms, which are
quite prominent in their own right.

However, a two-level atom interacting with a scalar field is more or
less a toy model, and a more realistic system would be a multilevel
atom, a hydrogen atom, for instance, in interaction with a quantized
electromagnetic field. Let us note that such a system in the
multi-polar coupling scheme was recently examined in terms of  the
spontaneous excitation of an accelerated atom in both a free
space~\cite{ZYL06} and cavities~\cite{YZ06}.  It has been found that
both the effects of vacuum fluctuations and radiation reaction on
the atom are altered by the acceleration. This differs from the
scalar field case where the contribution of radiation reaction is
not changed by the acceleration. A dramatic feature is that the
contribution of electromagnetic vacuum fluctuations to the
spontaneous excitation rate contains an extra term proportional to
$a^2$, the proper acceleration squared, in addition to the usual
Planckian thermal term of the Unruh temperature $ T=a/2\pi$. This is
in contrast to the scalar field case where the effect of
acceleration is purely thermal. As a further step along the line, in
this paper, we would like to study, using the DDC formalism,  the
spontaneous excitation of a static multi-level atom  in multi-polar
interaction with quantized electromagnetic fields in vacuum outside
a four dimensional Schwarzschild black hole. Our discussion will be
based upon the Gupta-Bleuler quantization of free electromagnetic
fields in a static spherically symmetric spacetime of arbitrary
dimension in a modified Feynman gauge given by Crispino et
al~\cite{LAG01}. At this point, let us note the DDC formalism has
also been applied to calculate the radiative energy shifts of
accelerated atoms both in a free space~\cite{Audretsch and
Muller,Audretsch95,Passante} and in
cavities~\cite{Rizzuto,Rizzuto09,ZY10}.

The paper is organized as follows, we introduce, in Sec. II, the
general formalism developed in Ref.~\cite{Audretsch94} and
generalized in Ref.~\cite{ZYL06} to the case of a multilevel atom
interacting with a quantized electromagnetic field in the multipolar
coupling scheme. In Sec. III, we first  review  the Gupta-Bleuler
quantization of free electromagnetic fields in the background
Schwarzschild black hole~\cite{LAG01}, and we then define, in
analogy to the scalar field case, the Boulware, Unruh and
Hartle-Hawking vacuum states, calculate the two-point functions for
the electromagnetic fields in these vacuum states and analyze their
properties in asymptotic regions. The calculation of the spontaneous
excitation of the multi-level atom interacting with a quantized
electromagnetic field in the multipolar coupling scheme in all
vacuum states will be performed in Sec. IV and a summary will be
given in Sect. V

\section{general formalism}

The DDC formalism is carried out in the Heisenberg picture that provides a
very convenient theoretical framework as it leads, for the relevant
dynamical variables, to equations of motion very similar to the corresponding
classical ones.

Consider a multilevel atom in interaction with vacuum
electromagnetic fluctuations outside a four-dimensional spherically
symmetric black hole. The line element of the spacetime is
 \beq
d^2s=(1-2M/r)\;d^2t-(1-2M/r)^{-1}d^2r
     -r^2(d\theta^2+\sin^2\theta d\phi^2)
 \eeq
with $M$ being the mass of the black hole. Let $x(\tau)$ represent
the stationary trajectory of the atom and $\tau$ denote its proper
time. The stationary trajectory condition guarantees the existence
of stationary states. By assuming the multi-polar coupling,
the total Hamiltonian that governs the evolution of the atom-field
system with regard to the proper time $\tau$ can be written as
 \beq
H(\tau)=H_A(\tau)+H_F(\tau)+H_I(\tau)\;. \label{total Hamiltonian}
 \eeq
 Here $H_A(\tau)$ is the Hamiltonian that determines the evolution
of the atom,  and it is given by
 \beq
H_A(\tau)=\sum_n \omega_n \sigma_{nn}(\tau)\;,\label{atomic Hamiltonian}
 \eeq
where $|n\rangle$ represents a complete set of stationary states of
the atom with energies $\omega_n$, and $\sigma_{nn}=|n\rangle\langle
n|$. $H_F(\tau)$ is the Hamiltonian that decides the evolution of
the free quantum electromagnetic field with respect to the proper time $\tau$,
 \beq
H_F(\tau)=\sum_{\vec{k}}\omega_{\vec{k}}a^{+}_{\vec{k}}a_{\vec{k}}\frac{dt}{d\tau}\;,
 \eeq
where $\vec{k}$ stands for the wave vector and polarization of the
field modes, $a^{+}_{\vec{k}}$ and $a_{\vec{k}}$ are the
annihilation and creation operators with momentum $\vec{k}$.
$H_I(\tau)$ describes the coupling between the multilevel atom and
the electromagnetic field. In the multipolar coupling
scheme~\cite{CPP95},
 \beq
H_I(\tau)=-e\;\mathbf{r}(\tau)\cdot\mathbf{E}(x(\tau))=
          -e\sum_{mn}\mathbf{r}_{mn}\cdot\mathbf{E}(x(\tau))\;\sigma_{mn}(\tau)\;,
 \eeq
where $e$ is the electron electric charge, $e\mathbf{r}$ the
electric dipole moment of the atom, $x(\tau)$ the atomic
spacetime coordinates and $\mathbf{E}(x(\tau))$ the electric
field operator of vacuum electromagnetic fields.

Generally,  atomic observables evolve with time as a result of
interaction between the atom and the field.  The rate of change of
an arbitrary observable, $O(\tau)$, of the atom is governed by the
Heisenberg equation
 \beq
\frac{d O(\tau)}{d\tau}=i[H_A(\tau),O(\tau)]-ie[\mathbf{r}(\tau)\cdot\mathbf{E}(x(\tau)),O(\tau)]\;.
 \eeq
Here we are interested in the second part on the right hand side of
the above equation that is due to the interaction between the atom
and the field
 \beq
\biggl(\frac{d O(\tau)}{d\tau}\biggr)_{coupling}=-ie[\mathbf{r}(\tau)\cdot\mathbf{E}(x(\tau)),O(\tau)]\;.
 \eeq
The field operator $\mathbf{E}(x(\tau))$ can be divided into two
parts as $\mathbf{E}=\mathbf{E}^f+\mathbf{E}^s$.  Here and after,
the operators with superscript $f$ represent the free parts that
exist even when there is no coupling between the atom and the field
and those with  superscript $s$ represent the source parts that are
induced by the interaction between them. However, a tricky issue
arises when we try to perform this decomposition in the above
equation, since $\mathbf{E}^f$ and $\mathbf{E}^s$ do not separately
commutate with the atomic observable. As a result, we can write
 \bea
\biggl(\frac{d O(\tau)}{d\tau}\biggr)_{coupling}&=&-ie(\lambda\mathbf{E}^f(x(\tau))\cdot[\mathbf{r}(\tau),O(\tau)]
 +(1-\lambda)[\mathbf{r}(\tau),O(\tau)]\cdot\mathbf{E}^f(x(\tau)))\nonumber\\
 &&-ie(\lambda\mathbf{E}^s(x(\tau))\cdot[\mathbf{r}(\tau),O(\tau)]
 +(1-\lambda)[\mathbf{r}(\tau),O(\tau)]\cdot\mathbf{E}^s(x(\tau)))\;.
 \eea
So, there exists an ambiguity of operator ordering. Dalibard,
Dupont-Roc and Cohen-Tannoudji (DDC)~\cite{DDC82,DDC84} proposed to
use a symmetric ordering ($\lambda=\frac{1}{2}$) so that the two
terms on the right hand side can be separately Hermitian and possess
an independent physical meaning that can be unambiguously identified
as the contributions of vacuum fluctuations (vf) and radiation
reaction(rr) respectively. Setting $O(\tau)$ to be the atomic energy
$H_A(\tau)$, we have
 \beq
\frac{d H_A(\tau)}{d\tau}=\biggl(\frac{d H_A(\tau)}{d\tau}\biggr)_{vf}+
                        \biggl(\frac{d H_A(\tau)}{d\tau}\biggr)_{rr}
 \eeq
with
 \beq
\biggl(\frac{d
H_A(\tau)}{d\tau}\biggr)_{vf}=-\frac{ie}{2}(\mathbf{E}^f(x(\tau))
                              \cdot[\mathbf{r}(\tau),\sum_n\omega_n\sigma_{nn}(\tau)]
                              +[\mathbf{r}(\tau),\sum_n\omega_n\sigma_{nn}(\tau)]
                              \cdot\mathbf{E}^f(x(\tau)))\;,\label{vf}\\
 \eeq
 \beq
\biggl(\frac{d
H_A(\tau)}{d\tau}\biggr)_{rr}=-\frac{ie}{2}(\mathbf{E}^s(x(\tau))
                              \cdot[\mathbf{r}(\tau),\sum_n\omega_n\sigma_{nn}(\tau)]
                              +[\mathbf{r}(\tau),\sum_n\omega_n\sigma_{nn}(\tau)]
                              \cdot\mathbf{E}^s(x(\tau)))\;.\label{rr}\\
 \eeq
Accordingly, we can also divide the dynamical variables of the atom,
$\sigma_{mn}$ or $\mathrm{r}_i(\tau)$,  into free and source parts,
\bea
&&\sigma_{mn}(\tau)=\sigma^f_{mn}(\tau)+\sigma^s_{mn}(\tau)\;,\\
&&\mathrm{r}_i(\tau)=\mathrm{r}^f_i(\tau)+\mathrm{r}^s_i(\tau)\;.
 \eea
Then solving perturbatively the following Heisenberg equations of
motion they satisfy
 \bea
&&\frac{d}{d\tau}\sigma_{mn}(\tau)=i(\omega_m-\omega_n)\sigma_{mn}(\tau)
  -ie\mathbf{E}(x(\tau))\cdot[\mathbf{r}(\tau),\sigma_{mn}(\tau)]\;,\label{eq of sigma}\\
&&\frac{d}{d\tau}\mathrm{r}_i(\tau)=i\sum_{mn}(\omega_m-\omega_n)(\mathrm{r}_i)_{mn}\sigma_{mn}(\tau)
 -ie[\mathbf{r}(\tau),r_i(\tau)]\cdot\mathbf{E}(x(\tau))\;,\label{eq of r}
 \eea
 to order $e$, we find
 \bea
&&\left\{
    \begin{array}{ll}
      \sigma^f_{mn}(\tau)=\sigma_{mn}^f(\tau_0)\;e^{i(\omega_m-\omega_n)(\tau-\tau_0)}\;,\\
      \sigma^s_{mn}(\tau)=-ie\int^{\tau}_{\tau_0}d\tau'\;\mathbf{E}^f(x(\tau'))
  \cdot[\mathbf{r}^f(\tau'),\sigma_{mn}^f(\tau)]\;,\label{sigma}
    \end{array}
  \right.\\
&&\left\{
    \begin{array}{ll}
    (\mathrm{r}_i^f(\tau))_{mn}=(\mathrm{r}_i^f(\tau_0))_{mn}e^{i(\omega_m-\omega_n)(\tau-\tau_0)}\;,   \\
    \mathrm{r}_i^s(\tau)=-ie\int^{\tau}_{\tau_0}d\tau'
[\mathrm{r}_j^f(\tau'),\mathrm{r}_i^f(\tau)]\mathrm{E}_j^f(x(\tau'))\;.\label{ri}
    \end{array}
  \right.
 \eea
Here the repeated subscript denotes the summation over all spatial
components and $ (\mathrm{r}_i^f(\tau))_{mn}=\langle
m|\mathrm{r}_i^f(\tau)|n\rangle$ . Similarly, using the Heisenberg
equation for the annihilation operator of the field,
 \beq
\frac{d}{d\tau}a_{\vec{k}}(t(\tau))=-i\omega_{\vec{k}}a_{\vec{k}}(t(\tau))
  -ie\mathbf{r}(\tau)\cdot[\mathbf{E}(x(\tau)),a_{\vec{k}}(t(\tau))]\frac{d\tau}{dt}\;.\\
 \eeq
one can show to the same order of perturbation that
 \bea
&&\left\{
    \begin{array}{ll}
      a^f_{\vec{k}}(t(\tau))=a^f_{\vec{k}}(t(\tau_0))\;e^{-i\omega_{\vec{k}}[t(\tau)-t(\tau_0)]}\;,\\
      a^s_{\vec{k}}(t(\tau))=-ie\int^{\tau}_{\tau_0}d\tau'\;\mathbf{r}^f(\tau')
  \cdot[\mathbf{E}^f(x(\tau')),a^f_{\vec{k}}(t(\tau))]\label{af and
  as}\;,
    \end{array}
  \right.
 \eea
 and therefore the source filed $\mathrm{E}^s$ that is generated by interaction between the atom and the free field can be
 expressed as
 \beq
\mathrm{E}_i^s(x(\tau))=-ie\int^{\tau}_{\tau_0}d\tau'
\mathrm{r}_j^f(\tau')[\mathrm{E}_j^f(x(\tau')),\mathrm{E}_i^f(x(\tau))]\;.
\label{es}
 \eeq

Now we suppose that  the field is a vacuum state and the atom is in
state $|b\rangle$. For simplicity, we also assume that the atom is
polarized along the radial direction defined by the position of the
atom relative to the black hole space-time rotational killing
fields. This assumption significantly simplifies, while keeping the
physical essence of the problem, the computations we are going to
perform since we do not need to calculate the contributions
associated with the polarizations in the $\theta-$ and $\phi-$
directions. Taking the expectation values of Eqs.~(\ref{vf}) and
(\ref{rr}) over the state of the system $|0,b\rangle$, we can, using
Eqs.~(\ref{sigma}), (\ref{ri}) and (\ref{es}),  show that the
contributions of vacuum fluctuations and radiation reaction to the
rate of change of the mean atomic energy are given, to order $e^2$,
by
 \bea
\biggl\langle\frac{dH_A(\tau)}{d\tau}\biggr\rangle_{vf}
&=&2ie^2\int_{\tau_0}^{\tau}d\tau'\;C^F(x(\tau),x(\tau'))\frac{d}{d\tau}
   \chi^A_b(\tau,\tau')\;,\label{general contribution of vf}\\
\biggl\langle\frac{dH_A(\tau)}{d\tau}\biggr\rangle_{rr}
&=&2ie^2\int_{\tau_0}^{\tau}d\tau'\;\chi^F(x(\tau),x(\tau'))\frac{d}{d\tau}
   C^A_b(\tau,\tau')\;.\label{general contribution of rr}
 \eea
In the above equations, $C^F$ and $\chi^F$ are the two statistical
functions of the electromagnetic field, i.e., the symmetric
correlation function and the linear susceptibility function.  The
radial components of these function, which are relevant in our
future calculations, are defined as
 \bea
C^F(x(\tau),x(\tau'))&=&\frac{1}{2}\langle
0|\{E^f_r(x(\tau)),E^f_r(x(\tau'))\}|0\rangle\;,\label{general
C^F}\\
\chi^F(x(\tau),x(\tau'))&=&\frac{1}{2}\langle
0|[E^f_r(x(\tau)),E^f_r(x(\tau'))]|0\rangle\;,\label{general chi^F}
 \eea
where $\{\;,\;\}$ denotes the anti-commutator and $|0\rangle$
represents the vacuum state of the field which will be defined in
the next Sect. Two statistical functions of the field are dependent
on the trajectory of the atom. Analogously, $C^A_b(\tau,\tau')$ and
$\chi^A_b(\tau,\tau')$ are  two atomic statistical functions which
are defined as
 \bea
C^A_b(\tau,\tau')&=&\frac{1}{2}\langle b|\{\mathrm{r}^f(\tau),\mathrm{r}^f(\tau')\}|b\rangle\;,\\
\chi^A_b(\tau,\tau')&=&\frac{1}{2}\langle
b|[\mathrm{r}^f(\tau),\mathrm{r}^f(\tau')]|b\rangle\;.
 \eea
They don't depend on the atomic trajectory and are determined only
by the internal structure of the atom itself. Their explicit forms
are given, with respect to $\tau$, by
 \bea
C^A_b(\tau,\tau')&=&\frac{1}{2}\sum_d|\langle
b|\mathrm{r}(0)|d\rangle|^2\;
[e^{i\omega_{bd}(\tau-\tau')}+e^{-i\omega_{bd}(\tau-\tau')}]\;,
\label{cf b}\\
\chi^A_b(\tau,\tau')&=&\frac{1}{2}\sum_d|\langle
b|\mathrm{r}(0)|d\rangle|^2\;[e^{i\omega_{bd}(\tau-\tau')}-
e^{-i\omega_{bd}(\tau-\tau')}]\;, \label{chi b}
 \eea
here $\omega_{bd}=\omega_b-\omega_d$ and the sum extends over the
complete set of atomic states.

Now it is clear that the calculation of the rate of change  of the
mean atomic energy  requires detailed knowledge on the quantization
of electromagnetic fields in the exterior region of the  black hole
and the specification of vacuum states. This is the main topic for
the next section.

\section{Quantization of electromagnetic fields in the exterior region of a
Schwarzschild black hole}

The quantization of the electromagnetic field in an static
spherically symmetric Schwarzschild-like spacetime has been carried
out by Crispino {\it et al} \cite{LAG01} using the Gupta-Bleuler
condition in a modified Feynmann gauge. Here we first give a brief
review of their basic results, and we then define the vacuum states
and calculate the statistical functions of the field. The Lagrangian
density of the electromagnetic field in a modified Feynman gauge is
 \beq
\mathcal{L}_F=\sqrt{-g}\biggl[-\frac{1}{4}F_{\mu\nu}F^{\mu\nu}-\frac{1}{2}\;G^2\biggr]
 \eeq
with $G=\nabla^{\mu}A_{\mu}+K^{\mu}A_{\mu}$ and $K^{\mu}$ being a
vector independent of $A_{\mu}$.  From the Lagrangian density, we
can write down the field equations.  If we choose $K^{\mu}$ to be
$K^{\mu}=(0,2M/r^2,0,0)$, the equation for $A_t$ decouples from
other ones~\cite{LAG01}. A complete set of solutions of the field
equations can then be denoted by $A^{(\lambda n;\omega lm)}_\mu$.
Here the label $"n"$ distinguishes between modes incoming from the
past null infinity $\mathcal{J}^{-}$ (denoted with $n=\leftarrow$)
and those going out from the past horizon $H^{-}$ (denoted with
$n=\rightarrow$). The modes with $\lambda=0$ are
 \beq
A^{(0 n;\;\omega
lm)}_{\mu}=(\;R^{(0n)}_{l}(\omega|r)\;Y_{lm}\;e^{-i\omega
t},0,0,0)\;,
 \eeq
where $Y_{lm}$ is the spherical harmonics and
$R^{(0n)}_{l}(\omega|r)$ satisfies the radial equation
 \beq
\biggl[\frac{\omega^2}{(1-2M/r)}+\frac{(1-2M/r)}{r^2}\frac{d}{dr}\biggl(r^2
    \frac{d}{dr}\biggr)-\frac{l(l+1)}{r^2}\biggr]R^{(0n)}_{l}(\omega|r)=0\;.
 \eeq
These modes are nonphysical as they do not satisfy the gauge
condition $G=0$ that are satisfied by all other modes with
$\lambda\neq0$. Modes with $\lambda=3$ are pure-gauge modes, and
they are given by
 \beq
A_{\mu}^{(3n;\omega lm)}=\nabla_{\mu}\Lambda^{(n\omega lm)}
 \eeq
with
 \beq
\Lambda^{(n\omega
lm)}=\frac{i}{\omega}\;R^{(0n)}_{l}(\omega|r)\;Y_{lm}\;e^{-i\omega
t}\;.
 \eeq
Those with $\lambda=1,2$ correspond to two classes of physical
modes. For the first class of physical modes ($\lambda=1$), $A_t=0$, and
 \beq
A_r^{(1n;\omega lm)}=R^{(1n)}_{l}(\omega |r)\;Y_{lm}\;e^{-i\omega t}
 \eeq
with $l\geq1$, where the radial function $R^{(1n)}_{l}(\omega |r)$
satisfies
 \beq
\frac{1}{r^2}\frac{d}{dr}\biggl[(1-2M/r)
    \frac{d}{dr}(r^2R^{(1n)}_{l}(\omega |r))\biggr]
    +\biggl[\frac{\omega^2}{(1-2M/r)}-\frac{l(l+1)}{r^2}\biggr]
    R^{(1n)}_{l}(\omega |r)=0\;.\label{radial equation0}
 \eeq
The angular components can  be expressed as
 \beq
A_i^{(1n;\;\omega
lm)}=\frac{1-2M/r}{l(l+1)}\frac{d}{dr}\biggl(r^2R^{(1n)}_{ l}(\omega
|r)\biggr)\partial_iY_{lm}\;e^{-i\omega t}\;.
 \eeq
For the second class of physical modes ($\lambda=2$), $A_t=A_r=0$,
 \beq
   A_i^{(2n;\omega
lm)}=R^{(2n)}_{l}(\omega |r)\;Y_i^{(lm)}\;e^{-i\omega t}
 \eeq
with $R^{(2n)}_{l}(\omega |r)$ obeying the following radial equation
 \beq
\biggl[\frac{\omega^2}{(1-2M/r)}+\frac{d}{dr}\biggl((1-2M/r)
    \frac{d}{dr}\biggr)
    -\frac{l(l+1)}{r^2}\biggr]
    R^{(2n)}_{l}(\omega |r)=0\label{radial equation1}
 \eeq
in which $Y_i^{(lm)}$ are the divergence-free vector spherical
harmonics on the unit 2-sphere satisfying
 \beq
\tilde{\nabla}^k(\tilde{\nabla}_k Y_i^{(lm)}-{\tilde{\nabla}}_i
Y_k^{(lm)})=-l(l+1)\;Y_i^{(lm)}\;,
 \eeq
 \beq
\int
d\Omega\;\tilde{\eta}^{ij}\;\overline{Y_i^{(lm)}}\;Y_j^{(l'm')}=\delta_{ll'}\delta_{mm'}\;.
 \eeq
Here the overline denotes the complex conjugation, $i$
angular variables on the unit 2-sphere $S^2$ with metric
$\tilde{\eta}_{ij}$ and inverse metric $\tilde{\eta}^{ij}$ with
signature $(+,+)$, $\tilde{\nabla}_i$  the associated covariant
derivative on $S^2$ and
$\tilde{\nabla}^i\equiv\tilde{\eta}^{ij}\tilde{\nabla}_j$. The above
four classes of modes form a complete set of basis for the quantum
electromagnetic field. The normalization of them are determined from
the canonical commutation relations of the fields by requiring
suitable commutation relations for the annihilation and creation
operators.

To quantize the electromagnetic field, let us define the general inner product as
 \beq
(A^{(\zeta)},A^{(\zeta')})\equiv\int_{\Sigma}d\Sigma_{\mu}\;W^{\mu}[A^{(\zeta)},A^{(\zeta')}]
 \eeq
in which
 \bea
W^{\mu}[A^{(\zeta)},A^{(\zeta')}]\equiv
i\;[\;\overline{A^{(\zeta)}_{\nu}}\;\Pi^{(\zeta')\mu\nu}
   -\overline{\Pi^{(\zeta)\mu\nu}}\;A^{(\zeta')}_{\nu}\;]\;,\\
\Pi^{\mu\nu}\equiv\frac{1}{\sqrt{-g}}\frac{\partial\mathcal{L}_F}
  {\partial[\nabla_{\mu}A_{\nu}]}=
-[F^{\mu\nu}+g^{\mu\nu}G]\quad\;\;
 \eea
and $d\Sigma_{\mu}=d\sigma\;n_{\mu}$, $A_{\mu}^{(\zeta)}\equiv
A_{\mu}^{(\lambda n;\omega lm)}$. The equal-time commutation
relations for the fields and their momentum operators are
 \beq
[\;\hat{A}_{\mu}(t,\mathbf{x}),\hat{A}_{\nu}(t,\mathbf{x'})\;]
=[\;\hat{\Pi}^{t\mu}(t,\mathbf{x}),\hat{\Pi}^{t\nu}(t,\mathbf{x'})\;]=0\;,\\
 \eeq
 \beq
[\;\hat{A}_{\mu}(t,\mathbf{x}),\hat{\Pi}^{t\nu}(t,\mathbf{x'})\;]
=\frac{i\delta^{\nu}_{\mu}}{\sqrt{-g}}\;\delta^{3}(\mathbf{x}-\mathbf{x}')\;,
 \eeq
where $\mathbf{x}$ and $\mathbf{x'}$ represent all spatial
coordinates. Expand the field operator in terms of the complete set
of basic modes as
 \beq
\hat{A}_\mu(t,\mathbf{x})=\sum_{\lambda
nlm}\int_0^\infty\frac{d\omega}{\sqrt{4\pi\omega}}\;
    [A^{(\lambda n;\omega lm)}_{\mu}(t,\mathbf{x})\;\hat{a}^{(\lambda n)}_{\omega lm}+
    \overline{A^{(\lambda n;\omega lm)}_{\mu}}(t,\mathbf{x})\;{\hat{a}}_{\omega lm}^{+(\lambda
n)}]\;,
 \eeq
By using the inner products of the field  and the commutation
relations between the field and momentum operators, the commutation
relations between the annihilation and creation operators are found
to be
 \bea
&&[\hat{a}^{(3n)}_{\omega lm},\hat{a}^{(3n')^+}_{\omega'
l'm'}]=-[\hat{a}^{(0n)}_{\omega lm},{\hat{a}^{(3n')^+}_{\omega'
l'm'}}]=\delta_{nn'}\delta_{ll'}\delta_{mm'}\delta{(\omega-\omega')}\;,\\
&&[\hat{a}^{(1n)}_{\omega lm},\hat{a}^{(1n')^+}_{\omega'
l'm'}]=\;\;\;[\hat{a}^{(2n)}_{\omega lm},{\hat{a}^{(2n')^+}_{\omega'
l'm'}}]=\delta_{nn'}\delta_{ll'}\delta_{mm'}\delta{(\omega-\omega')}\;
 \eea
and all other commutators vanish. The Gupta-Bleuler condition
\cite{Itzykson} requires that
 \beq
\hat{G}^+| phys \rangle=0\;,
 \eeq
where $\hat{G}^+$ is the positive-frequency part of
$\hat{G}=\nabla^{\mu}\hat{A}_{\mu}+K^{\mu}\hat{A}_{\mu}$ and
$|phys\rangle$ represents an arbitrary physical state. This is
equivalent to the following statements: the states obtained by
applying the creation operator $\hat{a}^{(3n)^+}_{\omega lm}$ are
all nonphysical; the physical states of the form
$\hat{a}^{(0n)^+}_{\omega lm}|phys\rangle$ have zero norm and are
orthogonal to any physical state.

Now we can define vacuum states and calculate the correlation
functions of the field.
We start with the Boulware vacuum state,
$|0\rangle_B$, which is defined by
 \beq
a^{(\lambda n)}_{\omega lm}|0\rangle_B=0\;,\;for\;all\;
(\lambda,n,\omega,l,m)\;with\;(\omega>0)\;.
 \eeq
In the computation of the correlation function ${_B}\langle
0|\hat{E}_r(t)\hat{E}_r(t')|0\rangle_B$ with
$\hat{E}_r=\hat{A}_{t;r}-\hat{A}_{r;t}$, the contributions of
nonphysical modes and pure-gauge modes are found to be canceled out,
and  only the contribution of the first class of physical modes is
left
 \bea
{_B}\langle
0|\hat{E}_r(x)\hat{E}_r(x')|0\rangle_{B}&=&\frac{1}{4\pi}\sum_{lm}
    \int_{0}^\infty d\omega\;\omega\;
    e^{-i\omega(t-t')}\;Y_{lm}(\theta,\varphi)\;Y^{\star}_{lm}(\theta',\varphi')
    \nonumber\\&&\quad\;\quad\;\;\times
    [\overrightarrow{R}_l(\omega|r)\overrightarrow{R}^{\star}_l(\omega|r')+
     \overleftarrow{R}_l(\omega|r)\overleftarrow{R}^{\star}_l(\omega|r')]\;.
     \label{Boulware concrete two point function}
 \eea
Here and after, for the sake of brevity, we omit the label
$\lambda=1$ as others will not appear in the correlation functions.
For the Unruh vacuum state, $|0\rangle_{U}$, similar to that in the
scalar field case \cite{Unruh76,J.Hartle76}, we define the positive
frequency for modes steming from  $H^-$, the past horizon of the
black hole, with respect to  the Killing vector $\xi=\partial_U$ and
those originating at infinity with respect to the Killing vector
$\eta=\partial_t$. For the Hartle-Hawking vacuum state,
$|0\rangle_{H}$, we define the incoming modes to be positive
frequency with respect to $\overline{v}$, the canonical affine
parameter on the future horizon, and the outgoing modes to be
positive frequency with respect to $\overline{u}$, the canonical
affine parameter on the past horizon. Repeat the same steps as that
in the  Boulware vacuum case, the correlation function satisfying
the corresponding boundary conditions in both the Unruh and
Hartle-Hawking vacuums are also found to be only associated with the
first class of physical modes. They are given respectively by
 \bea
{_U}\langle
0|\hat{E}_r(x)\hat{E}_r(x')|0\rangle_{U}&=&\frac{1}{4\pi}\sum_{lm}
    \int_{-\infty}^\infty d\omega\;\omega\;
    e^{-i\omega(t-t')}\;Y_{lm}(\theta,\varphi)\;Y^{\star}_{lm}(\theta',\varphi')
    \nonumber\\&&\quad\;\quad\;\;\times
    \biggl[\frac{\overrightarrow{R}_l(\omega|r)
      \overrightarrow{R}^{\star}_l(\omega|r')}{1-e^{-2\pi\omega/\kappa}}+
     \theta(\omega)\overleftarrow{R}_l(\omega|r)
      \overleftarrow{R}^{\star}_l(\omega|r')\biggr]\;,\quad\;\quad\;\quad\;
     \label{Unruh vacuum correlation function}
 \eea
and
 \bea
{_H}\langle
0|\hat{E}_r(x)\hat{E}_r(x')|0\rangle_{H}&=&\frac{1}{4\pi}\sum_{lm}
    \int_{-\infty}^\infty d\omega\;\omega\;
    \biggl[e^{-i\omega(t-t')}\;Y_{lm}(\theta,\varphi)\;Y^{\star}_{lm}(\theta',\varphi')
    \frac{\overrightarrow{R}_l(\omega|r)\overrightarrow{R}^{\star}_l(\omega|r')}
    {1-e^{-2\pi\omega/\kappa}}\nonumber\\&&\quad\;\quad\;\;+\;
    e^{i\omega(t-t')}\;Y^{\star}_{lm}(\theta,\varphi)\;Y_{lm}(\theta',\varphi')
    \frac{\overleftarrow{R}^{\star}_l(\omega|r)\overleftarrow{R}_l(\omega|r')}
    {e^{2\pi\omega/\kappa}-1}\biggr]\;,
    \label{Hartle-Hawking vacuum correlation function}
 \eea
where $\kappa=1/4M$ is the surface gravity of the black hole.
Details about the features of the  correlation functions calls for
the specific properties of the radial functions, so now we turn our
attention to the analysis of the radial functions. To do so, let us
further write the radial function as
 \beq
R^{(n)}_{l}(\omega
|r)=\frac{\sqrt{l(l+1)}}{\omega}\frac{\varphi^{(n)}_{\omega
l}(r)}{r^2}\;,
 \eeq
then Eq.~(\ref{radial equation0}) becomes
 \beq
\biggl[\frac{d^2}{dr_{\star}^2}+\omega^2-
   \biggl(1-\frac{2M}{r}\biggr)\frac{l(l+1)}{r^2}\biggr]
   \varphi^{(n)}_{\omega l}(r)=0\;,\label{general function of part R}
 \eeq
where $r_{\star}=r+2M\ln(r/2M-1)$ is the Regge-Wheeler tortoise
coordinate. It is difficult to exactly solve this equation in terms
of elementary functions, but fortunately,  two classes of physical
solutions in the two asymptotic regions single out
 \bea
&&\overrightarrow{\varphi}_{\omega l}(r)\sim
  \left\{
    \begin{array}{ll}
      e^{i\omega r_*}+\overrightarrow{\mathcal{R}_l}(\omega)\;e^{-i\omega r_*}\;,
             \quad\;\quad\;r\rightarrow 2M\;, \\
      \overrightarrow{\mathcal{T}_l}(\omega)\;e^{i\omega r_*},
             \quad\;\quad\;\quad\;\quad\;\quad\;\;r\rightarrow\infty\;,
    \end{array}
  \right.\label{asymptotic outgoing mode}\\
&&\overleftarrow{\varphi}_{\omega l}(r)\sim
 \left\{
    \begin{array}{ll}
      \overleftarrow{\mathcal{T}_l}(\omega)\;e^{-i\omega r_*}\;,
             \quad\;\quad\;\quad\;\quad\quad r\rightarrow2M\;, \\
      e^{-i\omega r_*}+\overleftarrow{\mathcal{R}_l}(\omega)\;e^{i\omega r_*}\;,
             \quad\;\quad\;r\rightarrow\infty\;.
    \end{array}
  \right.\label{asymptotic ingoing mode}
 \eea
Here $\mathcal{R}$ and $\mathcal{T}$ are, respectively, the
reflection and transmission coefficients and the following
relationships exist among them and  their complex conjugates,
 \bea
\left\{
  \begin{array}{ll}
    \overrightarrow{\mathcal{T}_l}(\omega)
        =\overleftarrow{\mathcal{T}_l}(\omega)=\mathcal{T}_l(\omega)\;, \\
    |\overrightarrow{\mathcal{R}_l}(\omega)|
        =|\overleftarrow{\mathcal{R}_l}(\omega)|\;, \\
    1-|\overrightarrow{\mathcal{R}_l}(\omega)|^2
        =1-|\overleftarrow{\mathcal{R}_l}(\omega)|^2=|\mathcal{T}_l(\omega)|^2\;,\\
    \overrightarrow{\mathcal{R}_l}^\star(\omega)\mathcal{T}_l(\omega)
        =-\mathcal{T}^\star_l(\omega)\overleftarrow{\mathcal{R}}_l(\omega)\;.
  \end{array}
\right.
 \eea
Although the exact forms for these coefficients demand an exact
solution of Eq.~(\ref{general function of part R}) which is a
formidable task, further studies (see Appendix) show that the
summation concerning the radial functions in the two asymptotic
regions, i.e., $r\rightarrow 2M$ and $r\rightarrow\infty$, behaves
as
 \bea
\sum_l(2l+1)\;|\overleftarrow{R}_l(\omega|r)|^2\sim
 \left\{
   \begin{array}{ll}
     \frac{\sum_{l}l(l+1)(2l+1)\;|\mathcal{T}_l(\omega)|^2}{(2M)^4\;\omega^2},
          \;\quad\;\quad\;r\rightarrow2M\;, \\
     \frac{8\omega^2}{3g_{00}^2}\;,\quad\;\quad\;\quad\;\quad\;\quad\;\quad\;\;
         \;\;\;\;r\rightarrow\infty\;,
   \end{array}
 \right.\label{asymptotic summation of ingoing modes}
 \eea
and
 \bea
\sum_l(2l+1)\;|\overrightarrow{R}_l(\omega|r)|^2\sim
 \left\{
   \begin{array}{ll}
     \frac{8\omega^2}{3g_{00}^2}+\frac{1}{6M^2g_{00}^2}\;,
      \quad\;\quad\;\quad\quad\;r\rightarrow2M\;, \\
     \frac{\sum_{l}l(l+1)(2l+1)\;|\mathcal{T}_l(\omega)|^2}{\omega^2r^4}\;,
      \quad\;\quad\;r\rightarrow\infty\;.
   \end{array}
 \right.\label{asymptotic summation of outgoing modes}
 \eea

\section{spontaneous excitation of an atom interacting with
vacuum electromagnetic field in Schwarzschild spacetime.}

Suppose that a static multilevel atom is held  at the radial
distance $r$ and it interacts with fluctuating quantum
electromagnetic fields in vacuum. Using the DDC formalism introduced
in section II and the results in the preceding section, we now
calculate the rate of change  of the mean atomic energy.

\paragraph{Boulware vacuum.}
In the Boulware vacuum state, the two statistical functions of the
electromagnetic field, i.e., the symmetric correlation function and
the linear susceptibility function, can be easily found from
Eq.~(\ref{Boulware concrete two point function}) to be
  \bea
C^F(x(\tau),x(\tau'))&=&\frac{1}{32\pi^2}\int_{0}^{\infty}d\omega\;\omega\;
   (e^{-\frac{i\omega\Delta\tau}{\sqrt{g_{00}}}}+e^{\frac{i\omega\Delta\tau}{\sqrt{g_{00}}}})
   \nonumber\\&&\quad\;\;\;\times
   \sum^{\infty}_{l=1}(2l+1)\;[\;|\overrightarrow{R}_l(\omega|r)|^2
   +|\overleftarrow{R}_l(\omega|r)|^2\;]\;,\label{Boulware cf}
 \eea
and
 \bea
\chi^F(x(\tau),x(\tau'))&=&\frac{1}{32\pi^2}\int_{0}^{\infty}d\omega\;\omega\;
   (e^{-\frac{i\omega\Delta\tau}{\sqrt{g_{00}}}}-e^{\frac{i\omega\Delta\tau}{\sqrt{g_{00}}}})
   \nonumber\\&&\quad\;\;\;\times
   \sum^{\infty}_{l=1}(2l+1)\;[\;|\overrightarrow{R}_l(\omega|r)|^2
   +|\overleftarrow{R}_l(\omega|r)|^2\;]\;.\label{Boulware chi}
 \eea
In obtaining the above results, we have used the relation $\Delta
\tau=\sqrt{g_{00}}\;\Delta t$ with $g_{00}=(1-2M/r)$ and the following
property of the spherical harmonics
 \beq
\sum^l_{m=-l}|Y_{lm}(\theta,\phi)|^2=\frac{2l+1}{4\pi}\;.
 \eeq
Inserting Eqs.~(\ref{Boulware cf}) and (\ref{chi b}) into
Eq.~(\ref{general contribution of vf}), we obtain the contribution
of vacuum electromagnetic fluctuations to the rate of change of the
mean atomic energy,
 \bea
\biggl\langle\frac{dH_A(\tau)}{d\tau}\biggr\rangle_{vf}&=&
-\frac{e^2g_{00}}{16\pi}\biggl[\sum_{\omega_b>\omega_d}|\langle
b|\mathrm{r}(0)|d\rangle|^2\;\omega_{bd}^2\;P(\omega_{bd},r)
\nonumber\\&&\quad\;\quad\;-\sum_{\omega_b<\omega_d}|\langle
b|\mathrm{r}(0)|d\rangle|^2\;\omega_{bd}^2\;P(-\omega_{bd},r)\biggr]\;,
\label{Boulware vf}
 \eea
where we have defined
 \beq
P(\omega,r)=\overrightarrow{P}(\omega,r)+\overleftarrow{P}(\omega,r)
 \eeq
with
 \bea
\overrightarrow{P}(\omega,r)=\sum_l(2l+1)\;|\overrightarrow{R}(\omega\sqrt{g_{00}}\;|r)|^2\;,\\
\overleftarrow{P}(\omega,r)=\sum_l(2l+1)\;|\overleftarrow{R}(\omega\sqrt{g_{00}}\;|r)|^2\;.
 \eea
Hereafter, the summation over $l$ is implied to range from $1$ to
$\infty$. Similarly, by inserting Eqs.~(\ref{Boulware chi}) and
(\ref{cf b}) into Eq.~(\ref{general contribution of rr}), the
contribution of radiation reaction to the rate of change of the mean
atomic energy is calculated out to be
 \bea
\biggl\langle\frac{dH_A(\tau)}{d\tau}\biggr\rangle_{rr}&=&
-\frac{e^2g_{00}}{16\pi}\biggl[\sum_{\omega_b>\omega_d}|\langle
b|\mathrm{r}(0)|d\rangle|^2\;\omega_{bd}^2\;P(\omega_{bd},r)
\nonumber\\&&\quad\;\quad\;+\sum_{\omega_b<\omega_d}|\langle
b|\mathrm{r}(0)|d\rangle|^2\;\omega_{bd}^2\;P(-\omega_{bd},r)\biggr]\;.
\label{Boulware rr}
 \eea
Adding up Eqs.~(\ref{Boulware vf}) and (\ref{Boulware rr}) yields
the total rate of change of the mean atomic energy
 \beq
\biggl\langle\frac{dH_A(\tau)}{d\tau}\biggr\rangle_{tot}=
-\frac{e^2g_{00}}{8\pi}\sum_{\omega_b>\omega_d}|\langle
b|\mathrm{r}(0)|d\rangle|^2\;\omega_{bd}^2\;P(\omega_{bd},r)\;.
\label{Boulware total0}
 \eeq
Here only the term with $\omega_b>\omega_d$ survives after the
addition and it is negative. This means that for an atom originally
in the ground state, the contributions of vacuum fluctuations and
radiation reaction to the total rate of change of the mean atomic
energy cancel, and, as a result, the ground state is stable.
However,  an atom originally in  an excited state can transition to
lower level states, since  the total rate of change of the mean
atomic energy is negative. Although, qualitatively, these features
are same as those for the atom in the Minkowski vacuum state in a
flat spacetime, the total rate of change also displays quantitative
difference which is embodied in the factor $P(\omega_{bd},r)$. A
comparison of Eq.~(\ref{Boulware total0}) with Eq.~(23) in
Ref.~\cite{YuLu05}, which gives the rate of change of the mean
atomic energy for an inertial atom in a flat space with a reflecting
boundary, shows that the two rates are quite similar, and the
appearance of $P(\omega_{bd},r)$ in Eq.~(\ref{Boulware total0}) can
be understood as a result of backscattering of the vacuum
electromagnetic field modes off the space-time curvature in much the
same way as the reflection of the field modes at the reflecting
boundary in a flat space-time. To further understand the effect of
backscattering caused by the spacetime curvature, let us now analyze
what occurs in two asymptotic regions, i.e., at the spatial infinity
and at the event horizon, which are regions of particular physical
interest to us.

At spatial infinity, i.e., $r\rightarrow\infty$, a further
simplification of the total rate of change by use of
Eqs.~(\ref{asymptotic summation of ingoing modes}) and
(\ref{asymptotic summation of outgoing modes}) yields
 \beq
\biggl\langle\frac{dH_A(\tau)}{d\tau}\biggr\rangle_{tot}\approx
-\frac{e^2}{3\pi}\sum_{\omega_b>\omega_d}|\langle
b|\mathrm{r}(0)|d\rangle|^2\;\omega_{bd}^4\;[1
+f(\omega_{bd},r)]\label{rate in Boulware infty}
 \eeq
in which
 \beq
f(\omega_{bd},r)=\frac{3\sum_ll(l+1)(2l+1)\;|
   \mathcal{T}_l(\omega_{bd}\sqrt{g_{00}})|^2}{8r^4\omega_{bd}^4}
 \eeq
is a grey-body factor that characterizes the backscattering of the
electromagnetic field modes off the spacetime curvature. Notice that
at infinity, $f(\omega_{bd},r)\sim0$, so the rate of change reduces
to that of an inertial atom in the Minkowski vacuum in flat
spacetimes with no boundaries, suggesting that the Boulware vacuum
at large radii is equivalent to the Minkowski vacuum.

When the atom is fixed near the event horizon, i.e., when
$r\rightarrow2M$, further simplification gives
 \beq
\biggl\langle\frac{dH_A(\tau)}{d\tau}\biggr\rangle_{tot}\approx
-\frac{e^2}{3\pi}\sum_{\omega_b>\omega_d}|\langle
b|\mathrm{r}(0)|d\rangle|^2\;\omega_{bd}^4\;\biggl[\biggl(1+\frac{a^2}{\omega_{bd}^2}\biggr)
+f(\omega_{bd},r)\biggr] \label{Boulware total 2M}
 \eeq
with
 \beq
a=\frac{M}{r^2\sqrt{g_{00}}}=\frac{M}{r^2\sqrt{1-2M/r}}\;,\label{acceleration}
 \eeq
where $a$ is the proper acceleration of the static atom. Notice the
appearance of an extra term proportional to $a^2$ here in contrast
to the scalar field case~\cite{ZhouYu07}. Let us note however that the presence of the terms proportional to the proper acceleration squared also appear in cases when the Unruh-DeWitt monopole detector is replaced by a dipole detector which couples  to the derivatives of a scalar field~\cite{ZhuYu07,Hinton83}.   The proper acceleration
$a$ diverges as the event horizon is approached, so does the rate of
change of the mean atomic energy. However, at infinity, where the
spacetime is asymptotically flat, $a\sim0$, so its contribution to
the rate is negligible. This result is in sharp contrast to that of
the scalar field case, where the rate of change is always
finite~\cite{ZhouYu07}. Except for the grey-body factor,
$f(\omega_{bd},r)$, here the result agrees with that of a
coaccelerated atom with a proper acceleration $a$ in interaction
with fluctuating electromagnetic fields in the Rindler
vacuum~\cite{ZZY} (the case with the temperature of the thermal bath
set to zero). The above discussions reveal that the Boulware vacuum
is the vacuum state of static observers outside a black hole, and it
resembles the Rindler vacuum in the flat spacetime since the static
atom is accelerating with the proper acceleration $a$ with respect
to locally free-falling observers.

\paragraph{Unruh vacuum.}
For a multilevel  atom in interaction with quantum electromagnetic
fluctuations in the Unruh vacuum, two statistical functions of the
field are easily obtained from Eq.~(\ref{Unruh vacuum correlation
function}) as follows
 \bea
C^F(x(\tau),x(\tau'))&=&\frac{1}{32\pi^2}\int_{-\infty}^{\infty}d\omega\;\omega\;
   \biggl(e^{-\frac{i\omega\Delta\tau}{\sqrt{g_{00}}}}+e^{\frac{i\omega\Delta\tau}{\sqrt{g_{00}}}}\biggr)
   \nonumber\\&&\quad\;\;\;\times
   \biggl[\frac{\sum_{l}(2l+1)\;|\overrightarrow{R}_l(\omega|r)|^2}{1-e^{-2\pi\omega/\kappa}}
   +\theta(\omega)\sum_{l}(2l+1)\;|\overleftarrow{R}_l(\omega|r)|^2\;\biggr]\;,
 \eea
 and
 \bea
\chi^F(x(\tau),x(\tau'))&=&\frac{1}{32\pi^2}\int_{-\infty}^{\infty}d\omega\;\omega\;
   \biggl(e^{-\frac{i\omega\Delta\tau}{\sqrt{g_{00}}}}-e^{\frac{i\omega\Delta\tau}{\sqrt{g_{00}}}}\biggr)
   \nonumber\\&&\quad\;\;\;\times
   \biggl[\frac{\sum_{l}(2l+1)\;|\overrightarrow{R}_l(\omega|r)|^2}{1-e^{-2\pi\omega/\kappa}}
   +\theta(\omega)\sum_{l}(2l+1)\;|\overleftarrow{R}_l(\omega|r)|^2\;\biggr]\;.
 \eea
Now the contributions of vacuum fluctuations and radiation reaction
to the rate of change of the mean atomic energy can be calculated by
inserting the statistical functions into Eqs.~(\ref{general
contribution of vf}) and (\ref{general contribution of rr}). For the
contribution of vacuum electromagnetic fluctuations, we have
 \bea
\biggl\langle\frac{dH_A(\tau)}{d\tau}\biggr\rangle_{vf}&=&-\frac{e^2g_{00}}{16\pi}
      \biggl\{\sum_{\omega_b>\omega_d}|\langle b|\mathrm{r}(0)|d\rangle|^2\;\omega^2_{bd}\times\nonumber\\&&\quad\quad\quad
      \biggl[\biggl(1+\frac{1}{e^{2\pi\omega_{bd}/\kappa_r}-1}\biggr)\overrightarrow{P}(\omega_{bd},r)
      +\frac{\overrightarrow{P}(-\omega_{bd},r)}{e^{2\pi\omega_{bd}/\kappa_r}-1}
      +\overleftarrow{P}(\omega_{bd},r)\biggr]\nonumber\\&&\quad\;\quad\;\;
      -\sum_{\omega_b<\omega_d}|\langle b|\mathrm{r}(0)|d\rangle|^2\;\omega^2_{bd}\times\nonumber\\&&\quad\quad\quad
   \biggl[\biggl(1+\frac{1}{e^{2\pi|\omega_{bd}|/\kappa_r}-1}\biggr)\overrightarrow{P}(-\omega_{bd},r)
                  +\frac{\overrightarrow{P}(\omega_{bd},r)}{e^{2\pi|\omega_{bd}|/\kappa_r}-1}
                  +\overleftarrow{P}(-\omega_{bd},r)\biggr]\biggr\}\;,\nonumber\\
 \eea
where we have defined
 \beq
\kappa_r=\frac{\kappa}{\sqrt{1-2M/r}}\;.
 \eeq
For the contribution of the radiation reaction,
 \bea
\biggl\langle\frac{dH_A(\tau)}{d\tau}\biggr\rangle_{rr}&=&-\frac{e^2g_{00}}{16\pi}
      \biggl\{\sum_{\omega_b>\omega_d}|\langle b|\mathrm{r}(0)|d\rangle|^2\;\omega^2_{bd}\times\nonumber\\&&\quad\quad\quad
            \biggl[\biggl(1+\frac{1}{e^{2\pi\omega_{bd}/\kappa_r}-1}\biggr)\overrightarrow{P}(\omega_{bd},r)
                  -\frac{\overrightarrow{P}(-\omega_{bd},r)}{e^{2\pi\omega_{bd}/\kappa_r}-1}+\overleftarrow{P}(\omega_{bd},r)\biggr]\nonumber\\
   &&\quad\;\quad\;\;+\sum_{\omega_b<\omega_d}|\langle b|\mathrm{r}(0)|d\rangle|^2\;\omega^2_{bd}\times\nonumber\\&&\quad\quad\quad
   \biggl[\biggl(1+\frac{1}{e^{2\pi|\omega_{bd}|/\kappa_r}-1}\biggr)\overrightarrow{P}(-\omega_{bd},r)
                  -\frac{\overrightarrow{P}(\omega_{bd},r)}{e^{2\pi|\omega_{bd}|/\kappa_r}-1}
                  +\overleftarrow{P}(-\omega_{bd},r)\biggr]\biggr\}\;.\nonumber\\
 \eea
Compared with the case of the Boulware vacuum, both the
contributions of vacuum electromagnetic fluctuations and radiation
reaction  are altered by  the appearance of a Planckian factor.
Adding them up, we obtain the total rate of change of the mean
atomic energy,
 \bea
\biggl\langle\frac{dH_A(\tau)}{d\tau}\biggr\rangle_{tot}&=&-\frac{e^2g_{00}}{8\pi}
      \biggl\{\sum_{\omega_b>\omega_d}|\langle b|\mathrm{r}(0)|d\rangle|^2\;\omega^2_{bd}
            \biggl[\biggl(1+\frac{1}{e^{2\pi\omega_{bd}/\kappa_r}-1}\biggr)\overrightarrow{P}(\omega_{bd},r)
                  +\overleftarrow{P}(\omega_{bd},r)\biggr]\nonumber\\
   &&\quad\;\quad\;\;-\sum_{\omega_b<\omega_d}|\langle
   b|\mathrm{r}(0)|d\rangle|^2\;\omega^2_{bd}\;
   \frac{\overrightarrow{P}(\omega_{bd},r)}{e^{2\pi|\omega_{bd}|/\kappa_r}-1}\biggr\}\;.
   \label{general totalrate of change}
 \eea
Now the delicate balance between vacuum fluctuations and radiation
reaction that ensures the stability of the atom in its ground state
in the Boulware vacuum no longer exists. The $\omega_b<\omega_d$
term which gives a positive contribution to the total rate of change
of the mean atomic energy makes the transition of the atom from the
ground state to an excited state possible, i.e., excitation
spontaneously occurs in the Unruh vacuum outside a black hole. The
structure of the rate of change also suggests that there is thermal
radiation from the black hole (represented by the Planckian term)
which is backscattered by spacetime curvature (represented by
$\overrightarrow P$).  It is this thermal radiation that renders the
spontaneous excitation possible. The temperature of the thermal
radiation is given by \beq
T=\frac{\kappa_r}{2\pi}=\frac{\kappa}{2\pi}\frac{1}{\sqrt{1-2M/r}}=(g_{00})^{-1/2}T_H\;,
 \eeq
with $T_H=\kappa/2\pi$ being the usual Hawking temperature of the
black hole. This is actually the  Tolman relation which gives the
temperature felt by a local observer.

 In order to gain more understanding, let us now examine
the behavior of the rate of change in two asymptotic regions. First,
when the atom is fixed near the event horizon, i.e., when
$r\rightarrow2M$, we can approximate,  by use of
Eqs.~(\ref{asymptotic summation of ingoing modes}) and
(\ref{asymptotic summation of outgoing modes}), the rate of change
as follows
 \bea
\biggl\langle\frac{dH_A(\tau)}{d\tau}\biggr\rangle_{tot}&\approx&-\frac{e^2}{3\pi}
      \biggl\{\sum_{\omega_b>\omega_d}|\langle b|\mathrm{r}(0)|d\rangle|^2\;\omega^4_{bd}
            \biggl[\biggl(1+\frac{1}{e^{\;\omega_{bd}/T}-1}\biggr)
            \biggl(1+\frac{a^2}{\omega_{bd}^2}\biggr)+f(\omega_{bd},r)\biggr]\nonumber\\
   &&\quad\quad-\sum_{\omega_b<\omega_d}|\langle
   b|\mathrm{r}(0)|d\rangle|^2\;\omega^4_{bd}\;
   \frac{1}{e^{ |\omega_{bd}|/T}-1}\biggl(1+\frac{a^2}{\omega_{bd}^2}\biggr)\biggr\}\;.
 \label{Unruh total 2M}
 \eea
A distinct feature in contrast  to the case of the scalar
fields~\cite{ZhouYu07} is the existence of an extra term
proportional to $a^2$, the proper acceleration squared. It is worth
pointing out here that the appearance of such a term has also been
found when one studies the spontaneous excitation of an accelerated
multilevel atom coupled with electromagnetic vacuum fluctuations in
a flat spacetime~\cite{ZYL06}. Noteworthily, one can show that,
close to the event horizon, $T\approx a/2\pi$ holds. This leads to a
remarkable observation, that is, the
 $\omega_b<\omega_d$
term, which  makes the spontaneous excitation possible, can be
viewed completely as a result of the Unruh effect due to the proper
acceleration that must exist to hold the atom static (refer to
Eq.~(29) in Ref.~\cite{ZYL06}). This demonstrates a complete equivalence between the effect of acceleration and
that of a gravitational field at the event horizon.

If the atom is placed far away from the black hole, i.e.,
$r\rightarrow\infty$, the total rate of change of the mean atomic
energy becomes
 \bea
\biggl\langle\frac{dH_A(\tau)}{d\tau}\biggr\rangle_{tot}&\approx&-\frac{e^2}{3\pi}
      \biggl\{\sum_{\omega_b>\omega_d}|\langle b|\mathrm{r}(0)|d\rangle|^2\;\omega^4_{bd}
            \biggl[1+\biggl(1+\frac{1}{e^{2\pi\omega_{bd}/\kappa_r}-1}
            \biggr)f(\omega_{bd},r)\biggr]\nonumber\\&&\quad\quad
             -\sum_{\omega_b<\omega_d}|\langle
             b|\mathrm{r}(0)|d\rangle|^2\;\omega^4_{bd}\;
   \frac{f(\omega_{bd},r)}{e^{2\pi|\omega_{bd}|/\kappa_r}-1}\biggr\}\;.
   \label{Unruh total infinity}
 \eea
Notice that thermal terms are all  multiplied by a grey-body factor,
$f(\omega_{bd},r)$, which vanishes at spatial infinity. The above
results are in accordance with the common belief that thermal flux
emanates from the black hole horizon and is backscattered and partly
depleted by the curved spacetime geometry on its way to infinity.
So,  as the atom is placed further and further away, the flux it
feels becomes weaker and weaker.

\paragraph{Hartle-Hawking vacuum.} Now let us look at the case in
which the electromagnetic fields are in the Hartle-Hawking vacuum
state.  Two statistical functions of the field can be easily found
from Eq.~(\ref{Hartle-Hawking vacuum correlation function}) to be
 \bea
C^F(x(\tau),x(\tau'))&=&\frac{1}{32\pi^2}\int_{-\infty}^{\infty}d\omega\;\omega\;
   \biggl(e^{-\frac{i\omega\Delta\tau}{\sqrt{g_{00}}}}+e^{\frac{i\omega\Delta\tau}{\sqrt{g_{00}}}}\biggr)
   \nonumber\\&&\quad\;\;\;\times
   \biggl[\frac{\sum_{l}(2l+1)\;|\overrightarrow{R}_l(\omega|r)|^2}{1-e^{-2\pi\omega/\kappa}}
   +\frac{\sum_{l}(2l+1)\;|\overleftarrow{R}_l(\omega|r)|^2}{e^{2\pi\omega/\kappa}-1}\;\biggr]\;,
 \eea
 and
 \bea
\chi^F(x(\tau),x(\tau'))&=&\frac{1}{32\pi^2}\int_{-\infty}^{\infty}d\omega\;\omega\;
   \biggl(e^{-\frac{i\omega\Delta\tau}{\sqrt{g_{00}}}}-e^{\frac{i\omega\Delta\tau}{\sqrt{g_{00}}}}\biggr)
   \nonumber\\&&\quad\;\;\;\times
   \biggl[\frac{\sum_{l}(2l+1)\;|\overrightarrow{R}_l(\omega|r)|^2}{1-e^{-2\pi\omega/\kappa}}
   -\frac{\sum_{l}(2l+1)\;|\overleftarrow{R}_l(\omega|r)|^2}{e^{2\pi\omega/\kappa}-1}\;\biggr]\;.
 \eea
Inserting them into Eqs.~(\ref{general contribution of vf}) and
(\ref{general contribution of rr}) yields the contributions of
vacuum fluctuations and radiation reaction to the rate of change of
the mean atomic energy
 \bea
\biggl\langle\frac{dH_A(\tau)}{d\tau}\biggr\rangle_{vf}&=&-\frac{e^2g_{00}}{16\pi}
      \biggl\{\sum_{\omega_b>\omega_d}|\langle b|\mathrm{r}(0)|d\rangle|^2\;\omega^2_{bd}
            \biggl[\frac{\overrightarrow{P}(-\omega_{bd},r)+\overleftarrow{P}(\omega_{bd},r)}{e^{2\pi\omega_{bd}/\kappa_r}-1}
            +\nonumber\\&&\quad\;\quad\;\quad\;\quad\;\;\;\biggl(1+\frac{1}{e^{2\pi\omega_{bd}/\kappa_r}-1}\biggr)
            \biggl(\overrightarrow{P}(\omega_{bd},r)+\overleftarrow{P}(-\omega_{bd},r)\biggr)\biggr]\nonumber\\
   &&\quad\;\quad\;\;-\sum_{\omega_b<\omega_d}|\langle b|\mathrm{r}(0)|d\rangle|^2\;\omega^2_{bd}
   \biggl[\frac{\overrightarrow{P}(\omega_{bd},r)+\overleftarrow{P}(-\omega_{bd},r)}{e^{2\pi|\omega_{bd}|/\kappa_r}-1}
            +\nonumber\\&&\quad\;\quad\;\quad\;\quad\;\;\;\biggl(1+\frac{1}{e^{2\pi|\omega_{bd}|/\kappa_r}-1}\biggr)
            \biggl(\overrightarrow{P}(-\omega_{bd},r)+\overleftarrow{P}(\omega_{bd},r)\biggr)\biggr]\biggr\}\;,
 \eea
 and
 \bea
\biggl\langle\frac{dH_A(\tau)}{d\tau}\biggr\rangle_{rr}&=&\frac{e^2g_{00}}{16\pi}
      \biggl\{\sum_{\omega_b>\omega_d}|\langle b|\mathrm{r}(0)|d\rangle|^2\;\omega^2_{bd}
            \biggl[\frac{\overrightarrow{P}(-\omega_{bd},r)+\overleftarrow{P}(\omega_{bd},r)}{e^{2\pi\omega_{bd}/\kappa_r}-1}
            -\nonumber\\&&\quad\;\quad\;\quad\;\quad\;\biggl(1+\frac{1}{e^{2\pi\omega_{bd}/\kappa_r}-1}\biggr)
            \biggl(\overrightarrow{P}(\omega_{bd},r)+\overleftarrow{P}(-\omega_{bd},r)\biggr)\biggr]\nonumber\\
   &&\quad\quad+\sum_{\omega_b<\omega_d}|\langle b|\mathrm{r}(0)|d\rangle|^2\;\omega^2_{bd}
   \biggl[\frac{\overrightarrow{P}(\omega_{bd},r)+\overleftarrow{P}(-\omega_{bd},r)}{e^{2\pi|\omega_{bd}|/\kappa_r}-1}
            -\nonumber\\&&\quad\;\quad\;\quad\;\quad\;\biggl(1+\frac{1}{e^{2\pi|\omega_{bd}|/\kappa_r}-1}\biggr)
            \biggl(\overrightarrow{P}(-\omega_{bd},r)+\overleftarrow{P}(\omega_{bd},r)\biggr)\biggr]\biggr\}\;.
 \eea
Then total rate of change readily follows
 \bea
\biggl\langle\frac{dH_A(\tau)}{d\tau}\biggr\rangle_{tot}&=&-\frac{e^2g_{00}}{8\pi}
    \biggl\{\sum_{\omega_b>\omega_d}|\langle
    b|\mathrm{r}(0)|d\rangle|^2\;\omega^2_{bd}\times\nonumber\\&&\quad\;\quad\;\quad\;\quad\;\quad\;
    \biggl(1+\frac{1}{e^{2\pi\omega_{bd}/\kappa_r}-1}\biggr)
    \biggl(\overrightarrow{P}(\omega_{bd},r)+\overleftarrow{P}(-\omega_{bd},r)\biggr)\nonumber\\&&
    \quad\;\quad\;\;-\sum_{\omega_b<\omega_d}|\langle b|\mathrm{r}(0)|d\rangle|^2\;\omega^2_{bd}
    \times\nonumber\\&&\quad\;\quad\;\quad\;\quad\;\quad\;
    \frac{1}{e^{2\pi\omega_{bd}/\kappa_r}-1}
    \biggl(\overrightarrow{P}(\omega_{bd},r)+\overleftarrow{P}(-\omega_{bd},r)\biggr)\biggr\}\;.
 \eea
As in the Unruh vacuum case, the positive $\omega_b<\omega_d$ term
also appears, and this term leads to spontaneous excitation of
static atoms in the Hartle-Hawking vacuum in the exterior region of
the black hole. Besides, the Planckian factor in the total rate of
change is a revelation of the thermal nature of the Hartle-Hawking
vacuum. To learn more, let us further study what happens in the
asymptotic regions.

When the atom is fixed at infinity, i.e., $r\rightarrow\infty$, the
rate of change of the mean atomic energy is
 \bea
\biggl\langle\frac{dH_A(\tau)}{d\tau}\biggr\rangle_{tot}
&\approx&-\frac{e^2}{3\pi}
      \biggl\{\sum_{\omega_b>\omega_d}|\langle b|\mathrm{r}(0)|d\rangle|^2\;\omega^4_{bd}
      \biggl(1+\frac{1}{e^{2\pi\omega_{bd}/\kappa_r}-1}\biggr)
      \biggl[1+f(\omega_{bd},r)\biggr]\nonumber\\
      &&\quad\;\;\;-\sum_{\omega_b<\omega_d}|\langle
      b|\mathrm{r}(0)|d\rangle|^2\;\omega^4_{bd}\;
      \frac{1}{e^{2\pi|\omega_{bd}|/\kappa_r}-1}
      \biggl[1+f(\omega_{bd},r)\biggr]\biggr\}\;,
 \eea
furthermore, at spatial infinity where $f(\omega_{bd},r)$ can be
taken as zero, the total rate of change is what one would obtain
when the atom is immersed in a thermal bath at the Hawking
temperature $T=T_H $ in a flat spacetime, while at the event
horizon, i.e., $r\rightarrow2M$,
 \bea
\biggl\langle\frac{dH_A(\tau)}{d\tau}\biggr\rangle_{tot}
&\approx&-\frac{e^2}{3\pi}
      \biggl\{\sum_{\omega_b>\omega_d}|\langle b|\mathrm{r}(0)|d\rangle|^2\;\omega^4_{bd}
      \biggl(1+\frac{1}{e^{2\pi\omega_{bd}/\kappa_r}-1}\biggr)
      \biggl[\biggl(1+\frac{a^2}{\omega_{bd}^2}\biggr)+f(\omega_{bd},r)\biggr]\nonumber\\
      &&\quad\;\;\;-\sum_{\omega_b<\omega_d}|\langle
      b|\mathrm{r}(0)|d\rangle|^2\;\omega^4_{bd}\;
      \frac{1}{e^{2\pi|\omega_{bd}|/\kappa_r}-1}
      \biggl[\biggl(1+\frac{a^2}{\omega_{bd}^2}\biggr)+f(\omega_{bd},r)\biggr]\biggr\}\;,
      \label{HH 2M}
 \eea
it is divergent as $a\rightarrow\infty$. In addition to the contribution
of the outgoing thermal radiation from the event horizon that exists in
the Unruh vacuum (refer to Eq.~(\ref{Unruh total 2M})), there is another
contribution to the total rate of change, the thermal term multiplied by
$f(\omega_{bd},r)$, which can be regarded as resulting from the incoming
thermal radiation from infinity. Both incoming and outgoing thermal radiation
are backscattered off the spacetime curvature on their way. This result
is consistent with our usual understanding that the Hartle-Hawking
vacuum is not empty at infinity but corresponds instead to a thermal
distribution of quanta at the Hawking temperature, thus it describes
a black hole in equilibrium with an infinite sea of black-body radiation.

A distinctive feature in contrast to the scalar field case is the
existence of the term proportional to $a^2$ that is nontrivial at
the event horizon and thus is physically important. Remarkably,
Eq.~(\ref{HH 2M}) is in structural similarity to the total rate of
change of the mean atomic energy of an uniformly accelerated atom
interacting with electromagnetic field fluctuations in a flat space
with a reflecting boundary (refer to Eq.~(60))~\cite{YZ06}),
reflecting again that the scattering of the electromagnetic field
modes off the spacetime curvature plays similar role as the
reflection of the field modes at boundaries in a flat spacetime.
This similarity is particularly striking at the event horizon, where
$T\simeq a/2\pi$.

\section{summary}

In summary, using the Gupta-Bleuler quantization of free
electromagnetic fields in a static spherically symmetric spacetime
of arbitrary dimension in a modified Feynman gauge given by Crispino
et al~\cite{LAG01},  we have defined, in analogy to the scalar field
case, the Boulware, Unruh and Hartle-Hawking vacuum states  outside
a four-dimensional Schwarzschild black hole, calculated the
two-point functions for the electromagnetic fields in these vacuum
states and analyzed their properties in asymptotic regions. We then
computed  the contributions of vacuum fluctuations and radiation
reaction to total rate of change of the mean energy for a radially
polarized static multilevel atom in interaction with quantum
electromagnetic fluctuations in all the three vacuum states.

Our results show that the static atoms in the ground state in the
Boulware vacuum are stable and this is in qualitative agreement with
the case where the atom is assumed to be in interaction with
quantized massless scalar fields~\cite{ZhouYu07}. However,
the spontaneous emission rate of the excited atoms close to the horizon
contains an extra term proportional to the squared proper
acceleration of the atom in contrast to the scalar field case, and
this rate is not well-behaved at the event horizon as a result of
the blow-up of the proper acceleration of the static atom with
respect to the free-falling local observers there (note that this
acceleration vanishes however at the spatial infinity). This is in
sharp contrast to that of the scalar field case, where the rate of
change of the mean atomic energy is always finite~\cite{ZhouYu07}.

For the static atoms in both the Unruh and the Hartle-Hawking
vacuums, the delicate balance between the contributions of vacuum
fluctuations and radiation reaction that ensures the stability of
the static atoms in ground state in the Boulware vacuum no longer
exists, so spontaneous excitation occurs in the exterior region of
the black hole. The spontaneous excitation rate of the static atoms
is in accordance with our usual understanding that the Unruh vacuum
describes a black hole with thermal radiation emitting from its
event horizon, whereas the Hartle-Hawking vacuum depicts a radiating
black hole in equilibrium with an infinite sea of black-body
radiation.

Distinctive features in contrast to the scalar field case are the
existence of the term proportional to the proper acceleration
squared in the rate of change of the mean atomic energy in  the
Unruh and the Hartle-Hawking vacuums and the structural similarity
in the spontaneous excitation rate between the static atoms outside
a black hole and uniformly accelerated atoms interacting with
electromagnetic field fluctuations in a flat space with a reflecting
boundary, which is particularly dramatic at the event horizon where
a complete equivalence exists.

\begin{acknowledgments}

This work was supported in part by the NSFC under Grants No. 11075083 and No. 10935013, the Zhejiang Provincial Natural Science Foundation of China under Grant No. Z6100077, the National Basic Research Program of China under Grant No. 2010CB832803, the Program for Changjiang Scholars and Innovative Research Team in University (PCSIRT,  No. IRT0964), the Hunan Provincial Natural Science Foundation of
China under Grant No. 11JJ7001, and Hunan Provincial Innovation Foundation For Postgraduate under Grant
No. CX2011B187.

\end{acknowledgments}

\appendix*
\section{on the asymptotic evaluation of mode sums}

Here, we derive the properties of the quantity
$\sum_l(2l+1)\;|R^{(n)}_l(\omega|r)|^2$ in two asymptotic regions. Let
us start with the incoming modes $\overleftarrow{R}_l(\omega|r)$, by
examining, at a fixed radial distance $r$, the correlation function
of the field in the Boulware vacuum, Eq.~(\ref{Boulware concrete two
point function})
 \bea
{_B}\langle 0|\hat{E}_r(\tau)\hat{E}_r(\tau')|0\rangle_B&=&
   \frac{1}{16\pi^2}\int_0^{\infty}d\omega\;\omega\sum_{l=1}^{\infty}(2l+1)\;
   [\;|\overrightarrow{R}_l(\omega|r)|^2+|\overleftarrow{R}_l(\omega|r)|^2\;]\;e^{-\frac{i\omega
   \Delta\tau}{\sqrt{g_{00}}}}\nonumber\\
   &\sim&\frac{1}{\pi^2(\Delta t)^4}\;.\label{general two point function}
 \eea
Here $\Delta t$ is the interval between the coordinate time and the
approximation is taken at spatial infinity, where the spacetime is
asymptotically flat.  So, the two point function in the Boulware
vacuum at large radii and that in the Minkowski vacuum should agree.
At spatial infinity ($r\rightarrow\infty$), it can be deduced from
Eq.~(\ref{asymptotic outgoing mode}) that
 \beq
\sum_l(2l+1)\;|\overrightarrow{R}_l(\omega|r)|^2\approx
     \frac{\sum_{l}l(l+1)(2l+1)\;|\mathcal{T}_l(\omega)|^2}{\omega^2r^4}\;.
 \eeq
This is very small as $r\rightarrow\infty$,  and thus can be
neglected in Eq.~(\ref{general two point function}).  Using
$\int_0^{\infty}\omega^3 e^{-i\omega x}=\frac{6}{x^4}$, we obtain
 \beq
\sum_l(2l+1)\;|\overleftarrow{R}_l(\omega|r)|^2\approx\frac{8\omega^2}{3g_{00}^2}\;.
 \eeq

The summation over the outgoing modes,
$\overrightarrow{R}_l(\omega|r)$, in the region $r\sim2M$ can be
obtained by solving the equation of the corresponding radial
function as follows. Let $\xi^2=r/2M-1$ and $q=4M\omega$, the radial
equation Eq.~(\ref{general function of part R}) can be simplified to
be
 \beq
\xi^2\frac{d^2\overrightarrow{\varphi}_l}{d\xi^2}+\xi\frac{d\overrightarrow{\varphi}_l}{d\xi}
+[\;q^2-(2l\xi)^2\;]\overrightarrow{\varphi}_l=0\;,
 \eeq
we have approximated $l(l+1)\xi^2$ by $(l\xi)^2$ since $\xi\sim0$.
The general solution of this equation can be expressed in terms of
the modified Bessel functions as
 \beq
{\overrightarrow{\varphi}_l}|_{r\rightarrow 2M}\sim a_{l}\;
K_{iq}(2l\xi)+b_l\;I_{-iq}(2l\xi)\;.\label{general solution near 2M}
 \eeq
To estimate the coefficients $a_l$ and $b_l$, let us look at the
radial equation Eq.~(\ref{general function of part R}) in the
limiting case, $l\rightarrow\infty$. Now the effective potential in
the equation is very large as compared to the other two terms for
fixed $r$ and $\omega$, therefore, one can deduce that
$\varphi_l\sim0$ for large $l$. As a result,  $b_l$ is an
exponentially small function of $l$ when  $l$ is large,  as
$I_{-iq}(2l\xi)\sim\frac{e^{l\xi}}{\sqrt{4\pi l\xi}}$ ($l\gg1)$. The
second part in Eq.~(\ref{general solution near 2M}) therefore makes
a bounded contribution to the summation of $\sum_{l}
(2l+1)\;|\overrightarrow{R}_l(\omega|r)|^2$ and it is negligible as
the  contribution of the first term is proportional to $\xi^{-4}$.
The coefficient  $a_l$ can then  be determined by comparing the
general solution, Eq.~(\ref{general solution near 2M}), with the
asymptotic solution
 \beq
\overrightarrow{\varphi}_{\omega l}(r)\sim
      e^{i\omega r_*}+\overrightarrow{\mathcal{R}_l}(\omega)\;e^{-i\omega
      r_*}\;,
 \eeq
and the result is
 \beq
a_l\sim\frac{2\;l^{-iq}\;e^{iq/2}}{\Gamma(iq)}\;.
 \eeq
Thus  the summation to the leading order is
 \bea
\sum_l(2l+1)\;|\overrightarrow{R}(\omega|r)|^2
&\sim&\frac{4}{\omega^2r^4\;\Gamma(iq)\;\Gamma(-iq)}
 \sum_l l(l+1)(2l+1)\;|K_{iq}(2l\xi)|^2\nonumber\\
&\approx&\frac{8\omega^2}{3g_{00}^2}+\frac{1}{6M^2g_{00}^2}\;.
 \eea
Here we have appealed to the trick of approximating the infinite
summation over $l$ by an integral.

The summation over the outgoing modes,
$\overrightarrow{R}_l(\omega|r)$, at infinity can be easily deduced
from the asymptotic solution of the outgoing mode,
Eq.~(\ref{asymptotic outgoing mode}) as
 \beq
\sum_l(2l+1)\;|\overrightarrow{R}_l(\omega|r)|^2\approx
     \frac{\sum_{l}l(l+1)(2l+1)\;|\mathcal{T}_l(\omega)|^2}{\omega^2r^4}\;.
 \eeq


\begin{thebibliography}{90}

\bibitem{Welton48} T. A. Welton, Phys. Rev. {\bf 74}, 1157 (1948).
\bibitem{GRF83} G. Compagno, R. Passante, and F. Persico, Phys. Lett.
         A {\bf 98}, 253 (1983).
\bibitem{Ackerhalt} J. R. Ackerhalt, P. L. Knight and J. H. Eberly,
         Phys. Rev. Lett. {\bf 30}, 456 (1973).
\bibitem{P.W.Milonni} P. W. Milonni, Phys. Script {\bf T21}, 102 (1988).
\bibitem{DDC82} J. Dalibard, J. Dupont-Roc and C. Cohen-Tannoudji,
         J. Phys. (France) {\bf 43}, 1617 (1982).
\bibitem{DDC84} J. Dalibard, J. Dupont-Roc and C. Cohen-Tannoudji,
         J. Phys. (France) {\bf 45}, 637 (1984).
\bibitem{YuLu05}H. Yu and S. Lu, Phys. Rev. D 72, 064022 (2005); 73,
         109901 (2006).
\bibitem{Audretsch94} J. Audretsch and R. M\"{u}ller,
         Phys. Rev. A {\bf 50}, 1755 (1994).
\bibitem{ZhouYu07} H. Yu and W. Zhou, Phys. Rev. D {\bf 76}, 044023
        (2007); {\it ibid} D {\bf 76}, 027503 (2007).
\bibitem{Unruh76} W. G. Unruh, Phys. Rev. D {\bf 14}, 870 (1976).
\bibitem{ZYL06} Z. Zhu, H. Yu and S. Lu, Phys. Rev. D {\bf 73}, 107501 (2006).
\bibitem{YZ06} H. Yu and Z. Zhu, Phys. Rev. D {\bf 74}, 044032 (2006).
\bibitem{LAG01} Luis C. B. Crispino, Atsushi Higuchi and George E. A.
         Matsas, Phys. Rev. D {\bf 63}, 124008 (2001).
\bibitem{Audretsch and Muller} J. Audretsch and R. M\"{u}ller, Phys.
         Rev. A {\bf 52}, 629 (1995).
\bibitem{Audretsch95} J. Audretsch, R. M\"{u}ller and
         M. Holzmann, Class. and Quantum Grav. {\bf 12}, 2927 (1995).
\bibitem{Passante} R. Passante, Phys. Rev. A {\bf 57}, 1590 (1998).
\bibitem{Rizzuto} L. Rizzuto, Phys. Rev. A {\bf 76}, 062114 (2007).
\bibitem{Rizzuto09} L. Rizzuto and S. Spagnolo, Phys. Rev. A {\bf 79}, 062110 (2009).
\bibitem{ZY10} Z. Zhu, H. Yu and Z. Zhu, Phys. Rev. A {\bf 82}, 042108 (2010).
\bibitem{CPP95}  G. Compagno,
R. Passante, and F. Persico, {\it Atom-Field Interactions and
Dressed Atoms} (Cambridge University Press, Cambridge 1995).
\bibitem{Itzykson} C. Itzykson and J. B. Zuber, Quantum Field Theory
        (McGrawHill, New York, 1980).
\bibitem{J.Hartle76} J. Hartle and S. Hawking, Phys. Rev. D {\bf 13},
        2188 (1976).
\bibitem{ZhuYu07} Y. Zhu, H. Yu and Z. Zhu , Class. and Quantum Grav. {\bf 24}, 95(2007).
\bibitem{Hinton83} K. J. Hinton, J. Phys. A  {\bf
16}, 1937 (1983).
\bibitem{ZZY} Z. Zhu and H. Yu, Chin. Phys. Lett. {\bf 25}, 1575 (2008).
\end{thebibliography}
\end{document}